\documentclass[final]{l4dc2020}

\usepackage{bm}
\usepackage{float}
\usepackage{mathtools}
\usepackage{tikz}\usetikzlibrary{arrows}
\usepackage{bbm}

\newcommand\copyrighttext{%
	\footnotesize \textbf{This paper was presented during the 2nd Annual Conference on Learning for Dynamics and Control.}}
\newcommand\copyrightnotice{%
	\begin{tikzpicture}[remember picture,overlay]
		\node[anchor=south,yshift=20pt] at (current page.south) {\fbox{\parbox{\dimexpr\textwidth-\fboxsep-\fboxrule\relax}{\copyrighttext}}};
	\end{tikzpicture}%
}

\newcounter{stepskim}
\newcommand{\stepkim}[1][\empty]% #1 = label (optional)
{\stepcounter{stepskim}%
 \par
 \hangindent=4em
 \hangafter=1
 \makebox[4em][l]{\textit{Step \arabic{stepskim}:}}%
 \ifx#1\empty\else #1 --\fi
}
{\setcounter{stepskim}{0}%
 \let\item=\step
 \parskip=0.25\baselineskip
 \parindent=0pt}%
{\par}

\newcommand{\RR}{\mathbb{R}}

\newcommand{\mNorm}[2]{\left\lVert {#2} \right\rVert_{#1}}

\newcommand{\mDefFunction}[3]{#1: #2 \rightarrow #3}

\newcommand{\mDef}{\coloneqq}
\newcommand{\mDistribution}[1]{\mathcal Q_{#1}}

\newcommand{\mExpectationSmall}[1]{\mathbb E [#1]}
\newcommand{\mExpectationExp}[2]{\mathbb{E}_{#1}\left[#2\right]}
\newcommand{\mExpectationExpSmall}[2]{\mathbb{E}_{#1}[#2]}
\newcommand{\mExpectationExpMed}[2]{\mathbb{E}_{#1}\Bigg[#2\Bigg]}

\newcommand{\mData}{\mathbb D}

\newcommand{\mOnes}[1]{\mathbbm 1_{#1}}

\newcommand{\mKdim}{\mathrm{dim}_K}
\newcommand{\mEdim}{\mathrm{dim}_E}

\newcommand{\mPa}{\theta}
\newcommand{\mV}[3]{V^{#1}_{{#2},{#3}}}
\newcommand{\mVplain}{V}
\newcommand{\mR}{\ell}
\newcommand{\mPo}[1]{\pi_{#1}}
\newcommand{\mRe}[1]{{\Delta}_{#1}}
\newcommand{\mtRe}[1]{\tilde {\Delta}_{#1}}
\newcommand{\mCre}{\mathrm{CR}}
\newcommand{\mBo}[3]{{\mathcal T}^{#1}_{#2,#3}}

\newcommand{\mMpcCost}[2]{J^#1_#2}

\newcommand{\mUpred}{u}
\newcommand{\mUpredOpt}{u^*}

\newcommand{\mXpred}{x}

\newcommand{\XX}{\mathbb X}
\newcommand{\UU}{\mathbb U}

\newcommand{\NN}{\mathbb {N}}

\newcommand{\OO}{\mathcal{O}}

\tikzstyle{block} = [draw, rectangle,
minimum height=1cm, minimum width=3cm, text width=2.5cm, align=center]
\tikzstyle{sum} = [draw, fill=blue!20, circle, node distance=1cm]
\tikzstyle{input} = [coordinate]
\tikzstyle{output} = [coordinate]
\tikzstyle{pinstyle} = [pin edge={to-,thin,black}]
\makeatletter
\newif\ifmygrid@coordinates
\tikzset{/mygrid/step line/.style={line width=0.80pt,draw=gray!80},
	/mygrid/steplet line/.style={line width=0.25pt,draw=gray!80}}
\pgfkeys{/mygrid/.cd,
	step/.store in=\mygrid@step,
	steplet/.store in=\mygrid@steplet,
	coordinates/.is if=mygrid@coordinates}
\def\mygrid@def@coordinates(#1,#2)(#3,#4){%
	\def\mygrid@xlo{#1}%
	\def\mygrid@xhi{#3}%
	\def\mygrid@ylo{#2}%
	\def\mygrid@yhi{#4}%
}
\newcommand\DrawGrid[3][]{%
	\pgfkeys{/mygrid/.cd,coordinates=true,step=1,steplet=0.2,#1}%
	\draw[/mygrid/steplet line] #2 grid[step=\mygrid@steplet] #3;
	\draw[/mygrid/step line] #2 grid[step=\mygrid@step] #3;
	\mygrid@def@coordinates#2#3%
	\ifmygrid@coordinates%
		\draw[/mygrid/step line]
		\foreach \xpos in {\mygrid@xlo,...,\mygrid@xhi} {%
				(\xpos,\mygrid@ylo) -- ++(0,-3pt)
				node[anchor=north] {$\xpos$}
			}
		\foreach \ypos in {\mygrid@ylo,...,\mygrid@yhi} {%
				(\mygrid@xlo,\ypos) -- ++(-3pt,0)
				node[anchor=east] {$\ypos$}
			};
	\fi%
}
\makeatother

\newtheorem{assumption}{Assumption}

\title[Bayesian model predictive control]{Bayesian model predictive control:
Efficient model exploration and regret bounds using posterior sampling}
\usepackage{times}
\author{%
 \Name{Kim P. Wabersich} \Email{wkim@ethz.ch}\\
 \addr Institute for Dynamic Systems and Control, ETH Zurich, Switzerland
 \AND
 \Name{Melanie N. Zeilinger} \Email{mzeilinger@ethz.ch}\\
 \addr Institute for Dynamic Systems and Control, ETH Zurich, Switzerland%
}
\begin{document}
\copyrightnotice
\maketitle

\begin{abstract}%
Tight performance specifications in combination with operational constraints make model predictive control (MPC) the method of choice in various industries. As the performance of an MPC controller depends on a sufficiently accurate objective and prediction model of the process, a significant effort in the MPC design procedure is dedicated to modeling and identification. Driven by the increasing amount of available system data and advances in the field of machine learning, data-driven MPC techniques have been developed to facilitate the MPC controller design. While these methods are able to leverage available data, they typically do not provide principled mechanisms to automatically trade off exploitation of available data and exploration to improve and update the objective and prediction model. To this end, we present a learning-based MPC formulation using posterior sampling techniques, which provides finite-time regret bounds on the learning performance while being simple to implement using off-the-shelf MPC software and algorithms. The performance analysis of the method is based on posterior sampling theory and its practical efficiency is illustrated using a numerical example of a highly nonlinear dynamical car-trailer system.
\end{abstract}

\begin{keywords}%
  Thompson sampling, posterior sampling , control, predictive control, regret bounds
\end{keywords}

\section{Introduction}\label{sec:introduction}
  %% This is the general situation and the 'big problem' in science
  Many autonomous systems of practical relevance such as
  autonomous cars, delivery drones, or chemical synthesis processes
  need to be optimally controlled using limited input authority subject to
  safety specifications in terms of state constraints.
  To meet these requirements, model predictive
  control (MPC) techniques have been developed since the 1970's. A
  distinct property of MPC is the possibility to ensure state and input
  constraint satisfaction in a principled way while providing
	approximately optimal control performance. As a result, MPC has enabled the
	development of a wide range of high performance control applications as discussed,
  e.g., in~\cite{Morari1999,Qin2000}.

  %% Narrower problem originating from the specific MPC mechanism
  The central mechanism of MPC is based on solving an
  open-loop optimal control problem, the MPC problem, at
	discrete time instances based on the current system state.
  More precisely, the future system evolution starting from the currently measured
  system state is simultaneously predicted and optimized in real time
  using a prediction model of the plant. Due to uncertainties in the prediction model and
  external disturbances, however, only the first element of the resulting optimal input
  sequence is applied to the system. At each time instance,
  the procedure of prediction and optimization is repeated, which
  introduces state feedback and therefore allows for disturbance compensation.
  Nevertheless, the resulting control performance	heavily relies
  on a sufficiently accurate prediction model of the underlying system dynamics,
  which typically results in time-consuming system modeling and
  identification procedures.
  In addition to the prediction model, the closed loop behavior is
  essentially determined through the objective function used
  in the MPC problem. While it is commonly assumed that the
  objective is given in closed form it may not be explicitly available,
  e.g., in case of complex or interactive applications. For example, the objective
  of a pick and place application could be determined by a person who provides feedback
  whether objects are placed correctly. The MPC objective function
  would then need to be infered from noisy samples.

  %% Yet narrower gap resulting from specific methods that try to avoid that problem
  To account for the modeling challenge, MPC approaches that are capable of leveraging learning-based prediction models
  have been investigated, see e.g. \cite{Hewing2018a,Carron2019,Kamthe2018,Koller2018,Soloperto2018},
  which generally assume availability of sufficiently informative system data.
  However, by \emph{passively} relying on available data, the resulting prediction model
  will only improve performance if the data is informative for the current task.
  Another option to generate sufficiently informative task-independent system
  data is to apply classical offline system identification procedures as
  described e.g. in \cite{Ljung1998}. While this enables sensible model
  estimation and even completely data-driven MPC controllers as
  proposed in~\cite{Yang2015,Coulson2019,Berberich2019},
  the main limitation is the high cost of obtaining
  informative data regardless of the control objective. Therefore,
  such approaches can become impractical, especially for nonlinear or
  high dimensional systems.
  
  % Dual MPC
  To balance between passive knowledge exploitation and
  objective-independent exploration, effective exploration-exploitation strategies
  have been developed in so-called dual-MPC approaches, see 
  e.g.~\cite{Mesbah2018} for an overview.
  Thereby the idea is to consider the potential advantage of obtaining
  relevant data in the future, e.g. through approximate stochastic dynamic
  programming as in~\cite{Hanssen2015,Klenske2016a,Heirung2017,Arcari2019}.
  While these techniques show promising results for simple tasks
  up to two state dimensions in combination with very short MPC
  prediction horizons, they are, so far, fundamentally limited to
  systems of low complexity.
  In addition, due to the rather crude approximation of the underlying
  stochastic dynamic programming problem, no theoretical 
  performance guarantees have been reported so far.
  
  %% Contributions
  The goal of this paper is to address these limitations for episodic learning tasks
  through a Bayesian learning-based MPC controller that automatically
  trades off exploration and exploitation while maintaining
  the computational complexity of conventional MPC.
  This is achieved by combining MPC with posterior sampling for reinforcement
  learning (RL) as originally proposed in \cite{Strens2000} and theoretically
  investigated by \cite{Osband2013} and \cite{Osband2014}.

  To trade off extraction of informative data and exploitation of already
  cumulated data, we propose a simple mechanism
  that samples an MPC controller at the beginning of each episode according
  to its posterior probability of being optimal with respect to the uncertain
  system dynamics and objective. Thereby, initial uncertainty about the
  optimal soft-constrained MPC controller leads to exploration of MPC controllers and
  generates explorative data collection in closed loop.
  As the posterior belief about the optimal soft-constrained MPC controller gets
  more certain through such explorative episodes, the MPC samples begin
  to aggregate around the optimal soft-constrained MPC controller for the plant,
  therefore automatically trading off exploration and exploitation.
  The resulting learning-based MPC controller yields a standard MPC
  problem and can be implemented using available algorithms and software
  packages, such as~\cite{Wang2010,Houska2011,Domahidi2012,forcesnlp2017}.
  The presented MPC allows for
  a rigorous, finite-time performance analysis with respect to
  the a-priori unknown optimal soft-constrained MPC controller for a specific
  system at hand by applying results from model-based RL,
  relating the degree of sub-optimality of a model-based
  controller to the respective model discrepancy that vanishes
  at a provable learning rate.

  %% Outline of the paper
  In the remainder of the paper we begin by formalizing the considered
  class of system dynamics and objective functions, provide the
  necessary background on MPC and state the
  formal problem formulation using the notion of Bayesian expected regret.
  Afterwards, the learning-based MPC scheme is presented, analyzed,
  and demonstrated using a highly nonlinear learning task.

\section{Model predictive control as an approximate optimal control policy}
  % System dynamics
  We consider discrete-time stochastic dynamical systems of the form
  \begin{align}\label{eq:system}
    x(k+1) = f(x(k), u(k); \mPa_f) + w(k), \quad k=0,1,2,..,N-1
  \end{align}
  with inputs $u(k)\in\RR^m$, states $x(k)\in\RR^n$,
  parameters $\mPa_f \in \RR^{n_{\mPa_f}}$, zero mean
  $\sigma_w$-sub-Gaussian process noise $w(k) \sim \mDistribution{w}$,
  and random initial condition $x(0)\sim\mDistribution{x(0)}$.
  % Constraints
  The system is subject to input constraints
  $u(k)\in\UU\subseteq\RR^m$ and state constraints
  $x(k)\in\XX\subseteq\RR^n$, which should be satisfied
  point-wise in time.
  % Objective
  The control objective is to minimize a
  time-varying stage cost function $\mDefFunction{\mR}{\NN \times \XX\times \UU}{\RR}$
  along system trajectories up to a finite time horizon $N$, i.e. minimizing
  \begin{align}\label{eq:objective}
    \mExpectationExp{W}{\sum_{k=0}^{N-1} \mR(k, x(k), u(k); \mPa_\mR)},
  \end{align}
  where $W \mDef [w(0), w(1), .., w(N-2)]$ and $\mPa_\mR\in\RR^{n_{\mPa_\mR}}$
  parametrizes the objective function.
  
  %% MPC as approximation
  While many relevant control problems can be stated in the above form,
  it is generally intractable to compute an optimal control policy that
  minimizes~\eqref{eq:objective}, which motivates the use of MPC techniques.
  % motivation of nominal MPC
  A simple yet efficient approximate control strategy
  to minimize \eqref{eq:objective} is based on repeatedly solving
  a constrained optimal control problem initialized at the currently
  measured state $x(k)$ in a shrinking horizon fashion.  While corresponding MPC
  formulations vary greatly in their complexity, the most simplistic
  formulation, sometimes referred to as nominal MPC, often
  provides sufficient practical properties in terms of performance and constraint satisfaction. 
  Thereby we optimize over a control sequence $\{\mUpred_{i|k}\}$
  subject to state and input constraints while neglecting zero
  mean additive disturbances. The resulting MPC problem is given by
  \begin{subequations}\label{eq:mpc_problem}
  \begin{align}
    \mMpcCost{\mPa}{k}(x) \mDef 
      \min_{\mUpred_{i|k}} &~~ c_\epsilon I(\epsilon) + \sum_{i=k}^{N-1} \mR(i, \mXpred_{i|k}, \mUpred_{i|k};\mPa_\mR)\label{eq:mpc_problem_cost} \\
      \text{s.t.} &~~ \mXpred_{k|k} = x, ~~ \epsilon\geq 0, \\
                  &~~ \mXpred_{i+1|k} = f(\mXpred_{i|k}, \mUpred_{i|k}; \mPa_f), ~ i = k,..,N-2, \\
                  &~~ \mXpred_{i|k} \in \XX(\epsilon_i), ~ i = k,..,N-1,\label{eq:mpc_problem_state_constraint} \\
                  &~~ \mUpred_{i|k} \in \UU, ~ i = k,..,N-1,
  \end{align}
  \end{subequations}
  where the additional cost term $c_\epsilon I(\epsilon)$ together
  with~\eqref{eq:mpc_problem_state_constraint}
  corresponds to a soft-constraint reformulation of the
  state constraints $\mXpred_{i|k} \in \XX$,
  accounting for the fact that closed loop system trajectories might differ from
  nominal predictions and therefore ensuring feasibility of \eqref{eq:mpc_problem}.
  For example, if $\XX = \{x \in\RR^n | g(x) \leq 0_{n_g}\}$ a soft-constraint reformulation
  can be obtained as $\XX(\epsilon_i) = \{x\in\RR^n | g(x) \leq \epsilon_i\mOnes{n_g}\}$
  with $\epsilon_i \geq 0$, and
  $I(\epsilon) = c_{1,\epsilon}^\top \epsilon + c_{2,\epsilon}\epsilon^\top \epsilon$
  \citep{Kerrigan2000}.
  
  %% Cost under optimal soft-constrained MPC defines the target where we want to end up
  In the following, we denote the expected closed loop cost-to-go at time $k$
  and state $x$ as
  \begin{align}\label{eq:optimal_cost}
    \mV{\mPa}{\tilde\mPa}{k}(x) \mDef
      \mExpectationExp{W}{
        \sum_{j=k}^{N-1} \mR(j, x(j), u(j);\mPa_\mR)
        ~\middle|~
        \begin{matrix*}[l]
          x(k) = x, \\ 
          u(j) = \mPo{\tilde \mPa}(j, x(j)), \\
          x(j+1) = f(x(j), u(j); \mPa_f) + w(j)
        \end{matrix*}
      }
  \end{align}
  with $\mPa \mDef (\mPa_\mR,\mPa_f)$ and $\mPo{\tilde\mPa}(j, x(j)) \mDef \mUpredOpt_{j|j}(x;\tilde\mPa)$,
  being the first element of the optimal input sequence of the MPC problem~\eqref{eq:mpc_problem}
  at time step $j$ with parameters $\tilde \mPa\mDef(\tilde\mPa_\mR,\tilde\mPa_f)$.

  \section{Problem formulation}

  % Unknown parameters and learning task
  We consider the case of unknown system dynamics
  and objective parametrization. More specifically, the
  learning problem is to improve the performance of the
  MPC controller $\mPo{\mPa}$ through data-based refinements
  of a-priori unknown parameters $\mPa$.
  % Episodic learning setting
  The efficient collection of system data with respect to
  the objective~\eqref{eq:objective} is carried out through
  repeated episodic interactions with the system~\eqref{eq:system}.
  % Episodic collection of system data
  During each episode $e=0,1,..,N_E-1$,  we need to
  provide a control policy that trades off information extraction
  and knowledge exploitation when applied to system \eqref{eq:system}
  at each sampling time step $k=0,1,..,N-1$.
  % Data
  The corresponding data, obtained up to $N_E$ episodes, is denoted by
  \begin{align}\label{eq:data}
    \mData_{N_E}\mDef
    \left\{
      \left(
        k,x_{k,e},u_{k,e},f(x_{k,e},u_{k,e};\mPa_f) + w_{k,e},\mR(k,x_{k,e},u_{k,e};\mPa_\mR)+\epsilon_{k,e}
      \right)_{k=0}^{N-1}
    \right\}_{e=0}^{N_E-1}
  \end{align}
  with zero mean $\sigma_\epsilon$-sub-Gaussian
  measurement noise $\epsilon_{k,e}$ on objective observations.
  % Prior knowledge
  Prior knowledge about the system parameters $\mPa$
  such as production or sensor tolerances of the
  plant to be controlled is considered to be given as
  $\mPa \sim \mDistribution{\mPa}$. In the following,
  $\mPa_e\sim\mDistribution{\mPa|\mData_e}$ denotes the
  posterior belief about $\mPa$ during episode $e$
  after data~\eqref{eq:data} has been observed.
  % $\mPa_f \sim \mDistribution{\mPa_f}$ and
  % $\mPa_\mR \sim \mDistribution{\mPa_\mR}$.
  The learning progress based on acquired data~\eqref{eq:data} after $N_E$ episodes
  is measured w.r.t. the Bayesian cumulative expected regret
  \begin{align}\label{eq:cumulative_regret}
    \mCre(N_E)\mDef
      \mExpectationExp{\mPa, \mPa_e, \mData_e}{ \sum_{e=0}^{N_E-1} \mRe{e} }
      \text{   with episodic regret   }
      \mRe{e} \mDef \mExpectationExp{x}{
      \mV{\mPa}{\mPa_e}{0}(x)
      -
      \mV{\mPa}{\mPa}{0}(x)
    }.
  \end{align}
  Here, the regret $\mRe{e}$ during each learning episode is taken with
  respect to the optimal soft-constrained MPC in terms of the nominal model
  and objective accuracy, that is, the MPC controller
  based on~\eqref{eq:mpc_problem} using the
  true system parameters $\mPa$.
\section{Bayesian model predictive control}
  \begin{minipage}[h]{0.55\linewidth}
  % Posterior sampling MPC
    The proposed learning-based MPC controller
    is based on posterior sampling as first proposed
    in the general reinforcement learning (RL)
    setting by~\cite{Strens2000}.
    % How to sample
    The resulting procedure is given in the Bayesian MPC Algorithm and
    works as follows. Based on the prior information $\mDistribution{\mPa}$
    about the system~\eqref{eq:system} and objective~\eqref{eq:objective}, a
    parameter realization $\mPa_e = (\mPa_{\mR,e}, \mPa_{f,e})$ is sampled
    from the posterior belief at the beginning of each episode. The sample
    parametrizes the MPC problem~\eqref{eq:mpc_problem} during the $e$-th
    episode, which results in an MPC controller that is sampled according
    to its a-posteriori probability of being optimal.
  \end{minipage}
  \hfill\vline\hfill
  \begin{minipage}{.4\linewidth}
    \textbf{Bayesian MPC Algorithm}
    \begin{algorithm}[H]
      \KwData{Parametric model $f$, $\mR$; Prior $\mDistribution{\mPa}$}
      Initialize $\mData_0 = \emptyset$\;

      \For{episodes $e = 0, 1, .., N_E$}{
        sample $\mPa_e \sim \mDistribution{\mPa|\mData_e}$\;

        % parametrize MPC policy $\mPo{\mPa_e}$\;

        \For{time steps $k = 0, 1, .., N-1$}{
          apply $u(k) = \mPo{\mPa_e}(k, x(k))$\;

          measure objective and state 
        }

        extend data set to obtain $\mData_{e+1}$
      }
    \end{algorithm}
  \end{minipage}
  %%

  % What sampling implies
  By applying the  sampled controller $u(k)=\mPo{\mPa_e}(k, x(k))$,
  we obtain measurements of the state evolution and
  the objective value, leading to an update of the data set to $\mData_{e+1}$
  after $N$ time steps. The collected data then
  refines the posterior belief about $\mPa$ and the process
  is repeated in the subsequent episode.
  This mechanism naturally causes exploration
  in case of large uncertainties in the posterior
  distribution $\mDistribution{\mPa|\mData_e}$ due to
  rich variation in system trajectories through diverse MPC
  controller samples. At the same time, this mechanism also exploits
  collected knowledge as the
  posterior belief starts to cumulate around a consistent
  model of the true system.
  As a consequence, the performance of the sampled MPC will
  converge to that of the nominally optimal soft-constrained MPC.

\section{Bound on finite-time learning performance}
  We apply the analysis provided by~\cite{Osband2013,Osband2014}
  to bound the cumulative regret~\eqref{eq:cumulative_regret}
  w.r.t. the nominal soft-constrained MPC controller using the true system dynamics.
  First, we reformulate the regret \eqref{eq:cumulative_regret}
  in terms of the optimal cost-to-go that corresponds to the sampled parameters
  $\mPa_e$ in episode $e$.
  This allows us in a second step to express the regret in
  terms of the learning progress of the system dynamics and
  objective function. By enforcing
  a regularity assumption on the expected cost-to-go under sampled MPC
  controllers we finally bound the regret in terms of
  posterior mean estimation errors of $f$ and $\mR$,
  allowing us to state the desired regret bound.

  For the instant regret in episode $e$ we have that
  \begin{align*}
    \mExpectationExp{\mPa, \mPa_{e}, x, \mData_e }{\mRe{e}}
      &= \mExpectationExp{\mPa, x, \mData_e}{
          \mExpectationExpMed{\mPa_{e}}{
            \underbrace{
              \mV{\mPa}{\mPa_e}{0}(x)
              }_{
                \text{Measured}
            }
            -
            \underbrace{
              \mV{\mPa}{\mPa}{0}(x)
              }_{
                \text{Unknown}
            }
            |
            \mPa, x, \mData_e  
          }
        }.
  \end{align*}
  Since $\mV{\mPa}{\mPa}{0}(x)$ is unknown, we
  instead consider the regret in terms of the sampled MPC
  controller applied to the corresponding sampled system,
  for which it is optimal:
  \begin{align}\label{eq:rotated_regret}
    \mExpectationExp{\mPa, \mPa_{e}, x, \mData_e }{\mtRe{e}}
      &= \mExpectationExp{\mPa, x, \mData_e}{
          \mExpectationExpMed{\mPa_{e}}{
            \underbrace{
              \mV{\mPa}{\mPa_e}{0}(x)
              }_{
                \text{Measured}
            }
            -
            \underbrace{
              \mV{\mPa_e}{\mPa_e}{0}(x)
              }_{
                \text{Known}
            }
            |
            \mPa, x, \mData_e
          }
        }.
  \end{align}
  Using standard Thompson sampling (posterior matching) arguments we can
  verify that
  \begin{align*}
    \mExpectationExp{\mPa, \mPa_{e}, x, \mData_e}{\mRe{e} - \mtRe{e}} =	
      \mExpectationExp{\mData_e, x}{
        \mExpectationExp{\mPa, \mPa_{e}}{
            \mRe{e} - \mtRe{e}|\mData_e, x
        }
      }
      \Rightarrow
      \mExpectationExp{\mPa, \mPa_{e}, x, \mData_e}{\mRe{e}} = \mExpectationExpSmall{\mPa, \mPa_{e}, x, \mData_e }{\mtRe{e}}
  \end{align*}
  holds, since $\mPa|\mData_e$ and $\mPa_e$ are equally distributed,
  yielding equally distributed MPC controllers $\mPo{\mPa|\mData_e}$,
  and $\mPo{\mPa_e}$, i.e. equally distributed time/state-to-input mappings,
  as well as equally distributed regrets\footnote{Formally this requires
  that $\mRe{e}$ is measurable with respect to the $\sigma$-algebra
  generated by the observed data $\mData_e$.} $\mRe{e}$ and $\mtRe{e}$ \citep{Russo2014}.

  Next, the goal is to express the regret
  $\mtRe{e}$ explicitly in terms of the
  posterior estimation accuracy of the functions
  $f$ and $\mR$ to be learned instead of the
  episodic cost difference~\eqref{eq:rotated_regret}.
  As introduced in \cite{Osband2013}, we define the recursive
  operator
  \begin{align}\label{eq:bellman_operator}
    \mBo{\mPa}{\tilde \mPa}{k} \mVplain(x)
      \mDef \mR(k, x, \mPo{\tilde \mPa}(k, x); \mPa)
      + \mExpectationExp{w}{\mVplain(x^+)|x^+=f(x, \mPo{\tilde\mPa}(k, x);\mPa)+w}
  \end{align} 
  at time steps $k = 0, 1, .., N-1$ to express the
  cost-to-go for system parameters $\mPa$ under an
  MPC controller using potentially different parameters $\tilde \mPa$. 
  The cost under the optimal soft-constrained MPC~\eqref{eq:optimal_cost} can 
  therefore be written as
  $\mV{\mPa}{\mPa}{k}(x)=\mBo{\mPa}{\mPa}{k}\mV{\mPa}{\mPa}{k+1}(x)$,
  in which case $\mBo{\mPa}{\mPa}{k}$ relates to the
  Bellman operator. Repeated application of this relation allows us to eliminate
  the term $\mV{\mPa}{\mPa_e}{0}(x)$ in~\eqref{eq:rotated_regret}.
  We sketch the corresponding derivation for the special
  case of zero process noise, i.e. $w(k)=0$, for which we expand
  $\mtRe{e}(x)$ recursively using~\eqref{eq:bellman_operator}:
  \begin{align*}
      & \mV{\mPa}{\mPa_e}{0}(x)- \mV{\mPa_e}{\mPa_e}{0}(x)
        = \mR(0, x, \mPo{\mPa_e}(0, x);\mPa)  + \mV{\mPa}{\mPa_e}{1}(f(x,\mPo{\mPa_e}(0, x);\mPa))
          - \mR(0, x, \mPo{\mPa_e}(0, x);\mPa_e) \\
        &\qquad  - \mV{\mPa_e}{\mPa_e}{1}(f(x,\mPo{\mPa_e}(0, x);\mPa_e))
          + \mV{\mPa_e}{\mPa_e}{1}(f(x,\mPo{\mPa_e}(0, x);\mPa))
          - \mV{\mPa_e}{\mPa_e}{1}(f(x,\mPo{\mPa_e}(0, x);\mPa))  \\
      &= \left(\mBo{\mPa}{\mPa_e}{0} 
          -\mBo{\mPa_e}{\mPa_e}{0}
          \right) \mV{\mPa_e}{\mPa_e}{1}(x)
          + \mV{\mPa}{\mPa_e}{1}(x(1))
          - \mV{\mPa_e}{\mPa_e}{1}(x(1)) \\
      &= \left(  \mBo{\mPa}{\mPa_e}{0}
          -\mBo{\mPa_e}{\mPa_e}{0}
          \right) \mV{\mPa_e}{\mPa_e}{1}(x)
          +
          \left(\mBo{\mPa}{\mPa_e}{1} 
          -\mBo{\mPa_e}{\mPa_e}{1}
          \right) \mV{\mPa_e}{\mPa_e}{2}(x(1))
          + \mV{\mPa}{\mPa_e}{2}(x(2))
          - \mV{\mPa_e}{\mPa_e}{2}(x(2)) \\
      &~~\vdots \\
      &= \sum_{k=0}^{N-1} \left(
          \mBo{\mPa}{\mPa_e}{k} - \mBo{\mPa_e}{\mPa_e}{k}
          \right)\mV{\mPa_e}{\mPa_e}{k+1}(x(k))
          + \underbrace{
              \mV{\mPa}{\mPa_e}{N}(x(N))
            }_{=0}
          - \underbrace{
              \mV{\mPa_e}{\mPa_e}{N}(x(N))
            }_{=0},
  \end{align*}
  where $x(0)=x$ and $x(k+1)=f(x(k), \mPo{\mPa_e}(k, x(k));\mPa)$. Including
  again the process noise $w(k)$, this result enables us to bound
  \small
  \begin{align}\nonumber
      \mExpectationSmall{\mtRe{e}}  \leq 
      &\mExpectationExp{ }{
          \sum_{k=0}^{N-1}
          \mExpectationExp{w(k)}{
            |  \mV{\mPa_e}{\mPa_e}{k+1}(f(x(k),u(k);\mPa)+w(k))
              - \mV{\mPa_e}{\mPa_e}{k+1}(f(x(k),u(k);\mPa_e)+w(k))|
          }
      }
      +\\\label{eq:two_rotated regret terms}
      &\mExpectationExp{}{
          \sum_{k=0}^{N-1}
          |  \mR(k, x(k), u(k);\mPa)-\mR(k, x(k), u(k);\mPa_e)|},
  \end{align}
  \normalsize
  where the outer expectation is taken w.r.t. $\mPa, \mPa_{e}, x, \mData_e$.
  Consequently, we can bound the second term in~\eqref{eq:two_rotated regret terms}
  in terms of
  $
    \mExpectationExpSmall{\mPa, x, \mData_e }{
      \mExpectationExpSmall{\mPa_{e}}{
          \sum_{k=0}^{N-1}
          |  \mR(k, x(k), u(k);\mPa) - \mR(k, x(k), u(k);\mPa_e)|
          ~ | ~ \mPa, x, \mData_e
      }
    },
  $
  that is, bounding the conditional posterior mean
  error of the cost $|\mR(.;\mPa)- \mR(.;\mPa_e)|$ based
  on the real problem parameter realization $\mPa$, initial
  condition $x$, and observed data $\mData_e$ up to episode $e$. 
  To derive a similar bound on the first term in
  \eqref{eq:two_rotated regret terms} with respect
  to the conditional posterior mean error of the dynamics,
  we follow \cite{Osband2014} and assume the following regularity
  property on the expected cost-to-go.
  \begin{assumption}\label{ass:continuity_v}
    For all $\mPa_e\in \RR^{n_\mPa}$ and $x^+,\tilde x^+\in\XX$ there
    exists a constant $L_V>0$ such that
    \begin{align*}
      \mExpectationExp{w(k)}{
            | \mV{\mPa_e}{\mPa_e}{k+1}( x^+ + w(k))
              - \mV{\mPa_e}{\mPa_e}{k+1}(\tilde x^+ + w(k)) |
          }
      \leq
      L_V \mNorm{2}{x^+ - \tilde x^+}.
    \end{align*}
  \end{assumption}
  Note that Assumption~\ref{ass:continuity_v} can, e.g., be satisfied for
  the standard case of linear dynamics and positive definite quadratic
  objectives. This allows us to bound the first term in
  \eqref{eq:two_rotated regret terms} in a similar fashion by\linebreak
  $
    \mExpectationExpSmall{\mPa, x, \mData_e }{
      \mExpectationExpSmall{\mPa_{e}}{
          \sum_{k=0}^{N-1}
          L_V\mNorm{2}{f(x(k),u(k);\mPa)-f(x(k),u(k);\mPa_e)}
          ~ | ~ \mPa, x, \mData_e
      }
    }.
  $
  
  The previously outlined analysis steps provide a
  bound on the expected regret in terms of the
  deviation between the posterior mean estimates of
  $f$ and $\mR$, conditioned on observed data,
  and the true underlying system dynamics and objective function.
  As first proposed by \cite{Russo2014} for the case of
  real-valued functions in the bandit optimization setting and
  later extended by \cite{Osband2014} to vector-valued functions
  in the context of RL, the magnitude of this deviation can be
  described using two distinct measures of complexity.
  The first measure is given by the classical Kolmogorov dimensions
  $\mKdim(\mR)$ and $\mKdim(f)$ describing the complexity of $\mR$
  and $f$ in the parameters $\mPa$, see, e.g., \citet[Section 7.1]{Russo2014}.
  The other measure is
  called Eluder dimension, denoted by $\mEdim(\mR)$ and $\mEdim(f)$
  and describes the complexity of the mean inference problem based
  on sequentially obtained measurements. More details
  and explicit bounds on $\mEdim(\mR)$ can be found
  in~\citet[Section 4.1]{Osband2014}.
  Using these measures of complexity, we get the
  following regret bound as an immediate consequence
  of \citet[Theorem 1]{Osband2014} with $\tilde \OO$
  neglecting terms that are logarithmic in $N_E$.
  \begin{corollary}\label{cor:general_regret_bound}
    Let Assumption~\ref{ass:continuity_v} hold. If
    there exist constants $c_\mR$ and $c_f$ such that
    for all admissible $x\in\RR^n$, $u\in\RR^m$, $\mPa\in\RR^{n_{\mPa_\mR}}$,
    and $k=0,1,..,N$ it holds
    $\mR(k,x,u;\mPa_\mR)\leq c_\mR$, and $f(x,u;\mPa_f)\leq c_f$, then
    it follows that
    \begin{align*}%\label{eq:cor_general_regret_bound}
      \mCre(N_E)
      \leq 
      \tilde \OO\left(
        \sigma_\epsilon\sqrt{\mKdim(\mR)\mEdim(\mR)N_E N}
        +L_V \sigma_w \sqrt{\mKdim(f)\mEdim(f)N_E N}
      \right).
    \end{align*}
  \end{corollary}
  As a direct consequence of \citet[Proposition~2]{Osband2014} we
  obtain the following bound for the important special case
  of linear Bayesian regression.
  \begin{corollary}\label{cor:specific_regret_bound}
    Let the assumptions of Corollary~\ref{cor:general_regret_bound} hold.
    If $f(x,u;\mPa_f) = \mPa_f^\top \Phi_f(x, u)$
    and $l(x,u;\mPa_f) = \mPa_\mR^\top \Phi_\mR(x, u)$
    with $\mPa_\mR\in\RR^{n_\mR}$ and $\mPa_f\in\RR^{n_f\times n}$,
    then
    \begin{align*}
      \mCre(N_E)
      \leq 
      \tilde \OO\left(
        \sigma_\epsilon\sqrt{n_\mR N_E N}
        +L_V \sigma_w n\sqrt{n n_f N_E N}
      \right).
    \end{align*}
  \end{corollary}
  % Meaning of this result
  While Corollary~\ref{cor:general_regret_bound} provides a general regret bound
  in terms of $\dim_E$ and $\dim_K$, Corollary \ref{cor:specific_regret_bound} ensures
  a finite-time learning progress through a sub-linear bound on the cumulated regret in case
  of linear Bayesian regression. However, also in the general case of
  Corollary~\ref{cor:general_regret_bound}, the regret bound scales naturally with
  the process and measurement noise, as well as with the regularity property of the
  expected cost-to-go according to Assumption~\ref{ass:continuity_v}.

  % \begin{remark}
    Note that the regret bounds are valid for the objective function~\eqref{eq:mpc_problem_cost}
    including the slack variables that indicate constraint violations. Consequently,
    the cumulative regret in this case also bounds the cumulated amount of expected
    constraint violation during different learning episodes.
  % \end{remark}
\section{Numerical results}
\begin{figure}[t]
\includegraphics[width=0.3\textwidth]{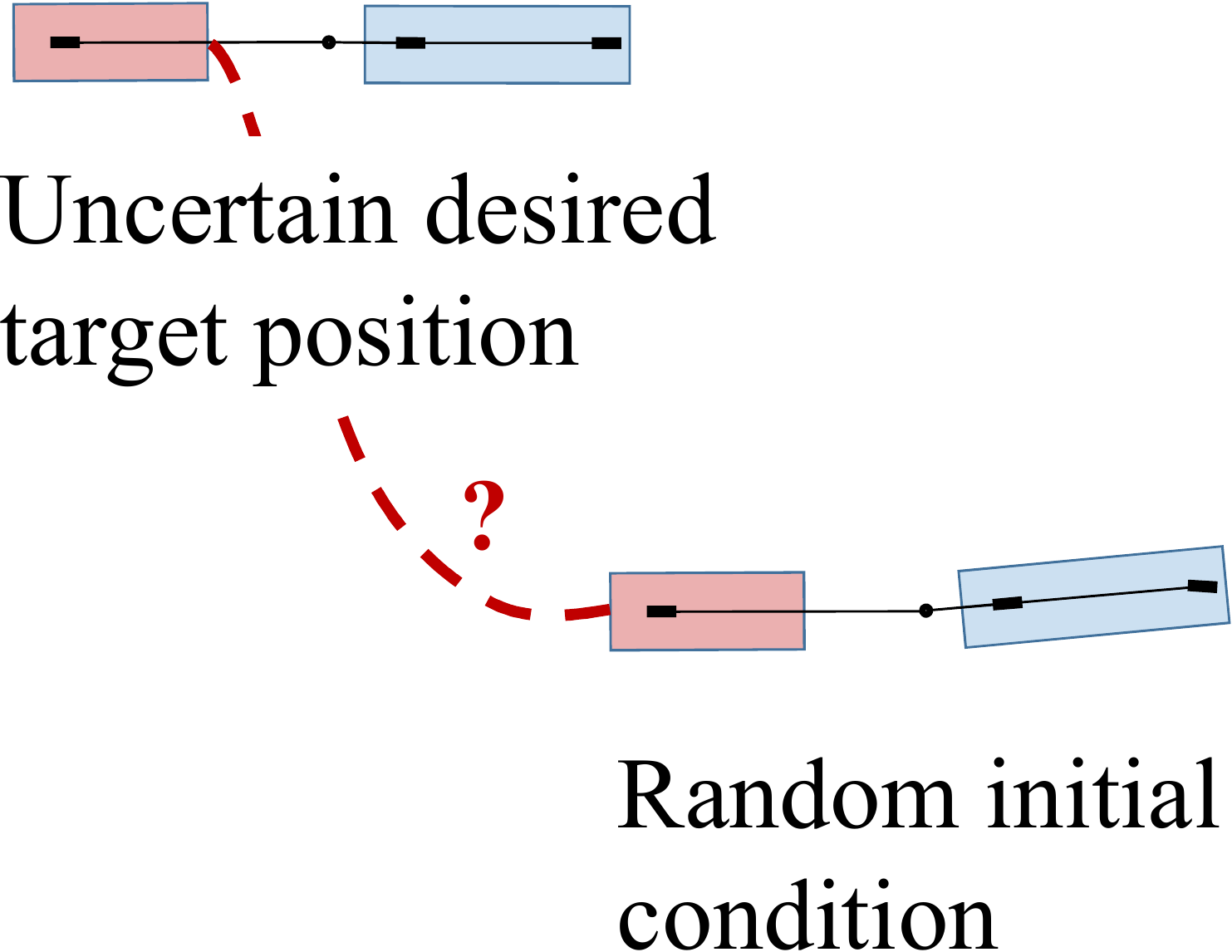}
\hfill
\vline
\hfill
\includegraphics[width=0.275\textwidth]{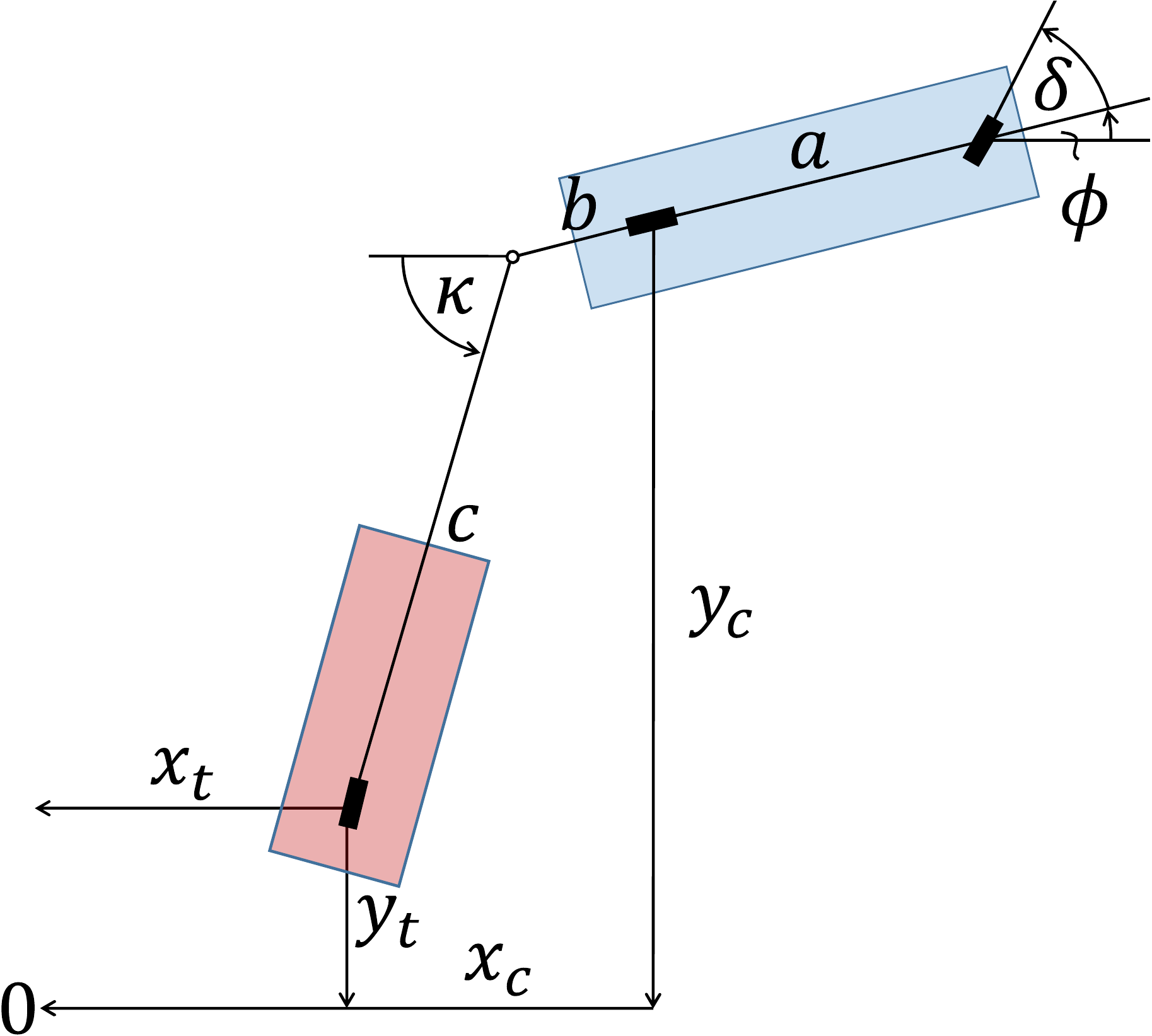}
\hfill
\vline 
\hfill 
\includegraphics[width=0.35\textwidth]{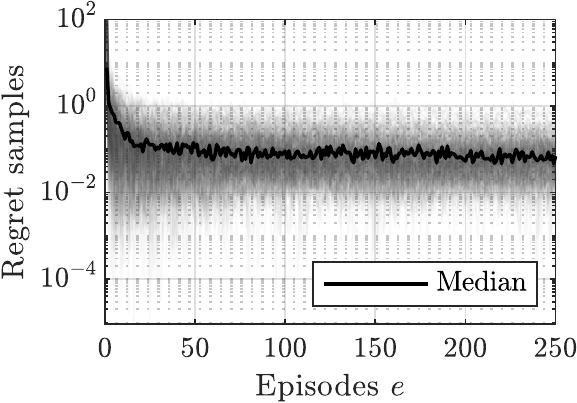}
\caption{\textbf{Left:} Problem setting of steering a partially known
  trailer system from a random initial condition to an uncertain target position.
  \textbf{Middle:} Car-trailer system. \textbf{Right:} Sampled regret
  $\Delta_e$ and median of 200 different systems over 250 learning episodes.}
\label{fig:car_trailer}
\end{figure}

% Example introduction
We consider the problem of learning how to drive a car-trailer system
with partially known dynamics backwards. Starting from a random
initial system configuration, the goal is to reach an uncertain goal
position as depicted in Figure~\ref{fig:car_trailer} (left) with
the car and trailer being horizontally aligned.
% Formal system description
The system dynamics according to Figure~\ref{fig:car_trailer} (middle)
are obtained through a Euler-forward discretization with sampling time $T_s = 0.1 \mathrm{[s]}$
of the model dynamics presented in~\cite{Rouchon1993}. By denoting $x^+ \mDef x(k+1)$,
$x\mDef x(k)$, and $u\mDef u(k)$ we get a prediction model of the form $x^+ = f(x, u;\mPa_f) + w$
with states $x=[y_c,\phi, \delta, \kappa, x_c, v_c]^\top$, inputs $u=[\omega_\delta, a_c]$, and
dynamics 
$$
\begin{array}{lll}
  x_c^+ = x_c + T_s v_c ~~~~& y_c^+ = y_c + T_s v_c \sin(\phi)
    ~~~~& v_c^+ = v_c + T_s a_c \\ 
  \phi^+ = \phi + T_s a^{-1}v\tan(\delta) & \delta^+ = \delta + \mPa_1^\top \Phi_1(x)
    & \kappa^+ = \kappa + \mPa_2^\top \Phi_2(x).
\end{array}
$$
The process noise $w\sim\mDistribution{w}$ accounts for model mismatch and
is given as $\mDistribution{w} = \NN(0, \Sigma_w)$ with
$\Sigma_w = \mathrm{diag}(T_s[0.03, 0.017, 0.1, .01, 0.01, 0.01])$. State and
input constraints are $|\kappa - \phi|\leq 0.7~\mathrm{[rad]}$, $x>1~\mathrm{[m]}$,  $|\delta| > 0.7~\mathrm{[rad]}$,
$|\omega_\delta|\leq 1.22~\mathrm{[rad]}$, and $|a|\leq 2\mathrm{[m/s^2]}$.
% Unknown terms in dynamics
The unknown terms in $\delta^+$ and $\kappa^+$ are parametrized by
features $\Phi_1(x) = \omega_\delta$ and $\Phi_2(x) =
  [v \sin(\kappa-\phi), v \tan(\delta) \cos(\kappa-\phi)]^\top$
and parameters $\mPa_1$ and $\mPa_2$,
describing the steering dynamics and the trailer geometry.
% Task objective
The objective of reaching the goal configuration can be encoded into a
terminal cost
$\mR(N-1, x, u)\mDef \phi^2 + \kappa^2 + v^2 + y_t^2 + x_t^2 + \mPa_\mR \Phi_3(x)$,
with $\mR(i, x,u)=0$ for $i=0,1,..N-2$,
where $\Phi_3(x)=[x_c^2, x_c, y_c^2 , y_c]^\top$ describes the desired target position. 
% Objective observation
After each episode, we obtain very noisy feedback from, e.g., a vision system or a person that is
modeled by zero mean normally distributed measurement noise $\epsilon_\mR\sim\NN(0, 0.5^2)$.

% Learning configuration
For learning, we consider a prior distribution $\mDistribution{\mPa}$ that corresponds
to a standard deviation of $0.45~\mathrm{[m]}$ in the trailer length, $10~\mathrm{[s/deg]}$
in steering dynamics, and $0.5~\mathrm{[m]}$ in the desired position $[x_d, y_d]^\top$.
% Remark on theoretical guarantees
Note that the theoretical results from Corollary~\ref{cor:general_regret_bound}
only hold, if the states and objectives are bounded within one episode, which is practically
fulfilled in this example due to the MPC controller that ensures
bounded input signals.

% Experiment execution
In Figure~\ref{fig:car_trailer} (right), we plot the measured
difference between the optimal and sampled MPC (=sampled regret), simulated
with 200 different system and objective realizations that are sampled according
to their prior distribution $\mDistribution{\mPa}$.
% Observations: Regret
During the first 15 episodes the median drops quickly to a slowly degrading regret,
depending on the process noise and measurement noise magnitude.

\section{Conclusion}
In this paper, we considered episodic learning
tasks for unknown dynamical systems and objective functions
subject to state and input constraints.
To enable efficient, easily implementable learning-based
control, we combined Bayesian posterior sampling theory
with model predictive control techniques. The learning
performance of the proposed approach can formally be bounded in terms of 
the regret w.r.t. the optimal model predictive controller. The efficiency of the algorithm
was demonstrated in simulation using a reverse driving task with a nonlinear car-trailer system.

% Acknowledgments---Will not appear in anonymized version
\acks{We thank Benjamin Van Roy for the fruitful discussion at ETH Zurich in spring 2017.
This work was supported by the Swiss National Science Foundation under grant no. PP00P2\_157601/1.}

\newpage

\bibliography{l4dc2020kim.bib}

\end{document}